\documentstyle[aps]{revtex}
\def\setRevDate $#1 #2 #3${#2}
\def\TeXdrawId{\setRevDate $Date: 1995/12/19 16:40:42 $ TeXdraw V2R0}

\chardef\catamp=\the\catcode`\@
\catcode`\@=11

\long                              % \centertexdraw needs to be very \long
\def\centertexdraw #1{\hbox to \hsize{\hss
                                      \btexdraw #1\etexdraw
                                      \hss}}

\def\btexdraw {\x@pix=0             \y@pix=0
               \x@segoffpix=\x@pix  \y@segoffpix=\y@pix
               \t@exdrawdef
               \setbox\t@xdbox=\vbox\bgroup\offinterlineskip
                   \global\d@bs=0           % pending segments
                   \global\t@extonlytrue    % no PS commands yet
                   \global\p@osinitfalse
                   \s@avemove \x@pix \y@pix % capture the initial position
                   \m@pendingfalse
                   \global\p@osinitfalse    % capture the next move
                   \p@athfalse
                   \the\everytexdraw}

\def\etexdraw {\ift@extonly \else
                 \t@drclose      % close the PostScript file
               \fi
               \egroup           % ends the \vbox \bgroup
               \ifdim \wd\t@xdbox>0pt
                 \t@xderror {TeXdraw box non-zero size,
                             possible extraneous text}%
               \fi
               \vbox {\offinterlineskip
                      \pixtobp \xminpix \l@lxbp  \pixtobp \yminpix \l@lybp
                      \pixtobp \xmaxpix \u@rxbp  \pixtobp \ymaxpix \u@rybp
                      \hbox{\t@xdinclude 
                        [{\l@lxbp},{\l@lybp}][{\u@rxbp},{\u@rybp}]{\p@sfile}}%
                      \pixtodim \xminpix \t@xpos  \pixtodim \yminpix \t@ypos
                      \kern \t@ypos
                      \hbox {\kern -\t@xpos
                             \box\t@xdbox       % TeX text
                             \kern \t@xpos}%
                      \kern -\t@ypos\relax}}

\def\drawdim #1 {\def\d@dim{#1\relax}}

\def\setunitscale #1 {\edef\u@nitsc{#1}%
                      \realmult \u@nitsc \s@egsc \d@sc}
\def\relunitscale #1 {\realmult {#1}\u@nitsc \u@nitsc
                      \realmult \u@nitsc \s@egsc \d@sc}
\def\setsegscale #1 {\edef\s@egsc {#1}%
                     \realmult \u@nitsc \s@egsc \d@sc}
\def\relsegscale #1 {\realmult {#1}\s@egsc \s@egsc
                     \realmult \u@nitsc \s@egsc \d@sc}

\def\bsegment {\ifp@ath
                 \f@lushbs
                 \f@lushmove
               \fi
               \begingroup
               \x@segoffpix=\x@pix
               \y@segoffpix=\y@pix
               \setsegscale 1
               \global\advance \d@bs by 1\relax}
\def\esegment {\endgroup
               \ifnum \d@bs=0
                 \writetx {es}%
               \else
                 \global\advance \d@bs by -1
               \fi}

\def\savecurrpos (#1 #2){\getsympos (#1 #2)\a@rgx\a@rgy
                         \s@etcsn \a@rgx {\the\x@pix}%
                         \s@etcsn \a@rgy {\the\y@pix}}
\def\savepos (#1 #2)(#3 #4){\getpos (#1 #2)\a@rgx\a@rgy
                            \coordtopix \a@rgx \t@pixa
                            \advance \t@pixa by \x@segoffpix
                            \coordtopix \a@rgy \t@pixb
                            \advance \t@pixb by \y@segoffpix
                            \getsympos (#3 #4)\a@rgx\a@rgy
                            \s@etcsn \a@rgx {\the\t@pixa}%
                            \s@etcsn \a@rgy {\the\t@pixb}}

\def\linewd #1 {\coordtopix {#1}\t@pixa
                \f@lushbs
                \writetx {\the\t@pixa\space sl}}
\def\setgray #1 {\f@lushbs
                 \writetx {#1 sg}}
\def\lpatt (#1){\listtopix (#1)\p@ixlist
                \f@lushbs
                \writetx {[\p@ixlist] sd}}

\def\lvec (#1 #2){\getpos (#1 #2)\a@rgx\a@rgy
                  \s@etpospix \a@rgx \a@rgy
                  \writeps {\the\x@pix\space \the\y@pix\space lv}}
\def\rlvec (#1 #2){\getpos (#1 #2)\a@rgx\a@rgy
                   \r@elpospix \a@rgx \a@rgy
                   \writeps {\the\x@pix\space \the\y@pix\space lv}}
\def\move (#1 #2){\getpos (#1 #2)\a@rgx\a@rgy
                  \s@etpospix \a@rgx \a@rgy
                  \s@avemove \x@pix \y@pix}
\def\rmove (#1 #2){\getpos (#1 #2)\a@rgx\a@rgy
                   \r@elpospix \a@rgx \a@rgy
                   \s@avemove \x@pix \y@pix}

\def\lcir r:#1 {\coordtopix {#1}\t@pixa
                \writeps {\the\t@pixa\space cr}%
                \r@elupd \t@pixa \t@pixa
                \r@elupd {-\t@pixa}{-\t@pixa}}
\def\fcir f:#1 r:#2 {\coordtopix {#2}\t@pixa
                     \writeps {\the\t@pixa\space #1 fc}%
                     \r@elupd \t@pixa \t@pixa
                     \r@elupd {-\t@pixa}{-\t@pixa}}
\def\lellip rx:#1 ry:#2 {\coordtopix {#1}\t@pixa
                     \coordtopix {#2}\t@pixb
                     \writeps {\the\t@pixa\space \the\t@pixb\space el}%
                     \r@elupd \t@pixa \t@pixb
                     \r@elupd {-\t@pixa}{-\t@pixb}}
\def\fellip f:#1 rx:#2 ry:#3 {\coordtopix {#2}\t@pixa
                     \coordtopix {#3}\t@pixb
                     \writeps {\the\t@pixa\space \the\t@pixb\space #1 fe}%
                     \r@elupd \t@pixa \t@pixb
                     \r@elupd {-\t@pixa}{-\t@pixb}}
\def\larc r:#1 sd:#2 ed:#3 {\coordtopix {#1}\t@pixa
                            \writeps {\the\t@pixa\space #2 #3 ar}}

\def\ifill f:#1 {\writeps {#1 fl}}     % Fill only
\def\lfill f:#1 {\writeps {#1 fp}}     % Stroke and fill

\def\htext #1{\def\testit {#1}%
              \ifx \testit\l@paren
                \let\next=\h@move
              \else
                \let\next=\h@text
              \fi
              \next {#1}}

\def\rtext td:#1 #2{\def\testit {#2}%
                    \ifx \testit\l@paren
                      \let\next=\r@move
                    \else
                      \let\next=\r@text
                    \fi
                    \next td:#1 {#2}}

\def\textref h:#1 v:#2 {\ifx #1R%
                          \edef\l@stuff {\hss}\edef\r@stuff {}%
                        \else
                          \ifx #1C%
                            \edef\l@stuff {\hss}\edef\r@stuff {\hss}%
                          \else  % default L
                            \edef\l@stuff {}\edef\r@stuff {\hss}%
                          \fi
                        \fi
                        \ifx #2T%
                          \edef\t@stuff {}\edef\b@stuff {\vss}%
                        \else
                          \ifx #2C%
                            \edef\t@stuff {\vss}\edef\b@stuff {\vss}%
                          \else  % default B
                            \edef\t@stuff {\vss}\edef\b@stuff {}%
                          \fi
                        \fi}

\def\avec (#1 #2){\getpos (#1 #2)\a@rgx\a@rgy
                  \s@etpospix \a@rgx \a@rgy
                  \writeps {\the\x@pix\space \the\y@pix\space (\a@type) %
                            \the\a@lenpix\space \the\a@widpix\space av}}

\def\ravec (#1 #2){\getpos (#1 #2)\a@rgx\a@rgy
                   \r@elpospix \a@rgx \a@rgy
                   \writeps {\the\x@pix\space \the\y@pix\space (\a@type) %
                             \the\a@lenpix\space \the\a@widpix\space av}}

\def\arrowheadsize l:#1 w:#2 {\coordtopix{#1}\a@lenpix
                              \coordtopix{#2}\a@widpix}
\def\arrowheadtype t:#1 {\edef\a@type{#1}}

\def\clvec (#1 #2)(#3 #4)(#5 #6)%
           {\getpos (#1 #2)\a@rgx\a@rgy
            \coordtopix \a@rgx\t@pixa
            \advance \t@pixa by \x@segoffpix
            \coordtopix \a@rgy\t@pixb
            \advance \t@pixb by \y@segoffpix
            \getpos (#3 #4)\a@rgx\a@rgy
            \coordtopix \a@rgx\t@pixc
            \advance \t@pixc by \x@segoffpix
            \coordtopix \a@rgy\t@pixd
            \advance \t@pixd by \y@segoffpix
            \getpos (#5 #6)\a@rgx\a@rgy
            \s@etpospix \a@rgx \a@rgy
            \writeps {\the\t@pixa\space \the\t@pixb\space
                      \the\t@pixc\space \the\t@pixd\space
                      \the\x@pix\space \the\y@pix\space cv}}

\def\drawbb {\bsegment
               \drawdim bp
               \linewd 0.24       % line width 1/300 inch = 0.24 bp
               \setunitscale {\p@sfactor}
               \writeps {\the\xminpix\space \the\yminpix\space mv}%
               \writeps {\the\xminpix\space \the\ymaxpix\space lv}%
               \writeps {\the\xmaxpix\space \the\ymaxpix\space lv}%
               \writeps {\the\xmaxpix\space \the\yminpix\space lv}%
               \writeps {\the\xminpix\space \the\yminpix\space lv}%
             \esegment}

\def\getpos (#1 #2)#3#4{\g@etargxy #1 #2 {} \\#3#4%
                        \c@heckast #3%
                        \ifa@st
                          \g@etsympix #3\t@pixa
                          \advance \t@pixa by -\x@segoffpix
                          \pixtocoord \t@pixa #3%
                        \fi
                        \c@heckast #4%
                        \ifa@st
                          \g@etsympix #4\t@pixa
                          \advance \t@pixa by -\y@segoffpix
                          \pixtocoord \t@pixa #4%
                        \fi}

\def\getsympos (#1 #2)#3#4{\g@etargxy #1 #2 {} \\#3#4%
                     \c@heckast #3%
                     \ifa@st \else
                       \t@xderror {TeXdraw: invalid symbolic coordinate}%
                     \fi
                     \c@heckast #4%
                     \ifa@st \else
                       \t@xderror {TeXdraw: invalid symbolic coordinate}%
                     \fi}

\def\listtopix (#1)#2{\def #2{}%
                      \edef\l@ist {#1 }%    % append a blank to the string
                      \m@oretrue
                      \loop
                        \expandafter\g@etitem \l@ist \\\a@rgx\l@ist
                        \a@pppix \a@rgx #2%
                        \ifx \l@ist\empty
                          \m@orefalse
                        \fi
                      \ifm@ore
                      \repeat}

\def\realmult #1#2#3{\dimen0=#1pt
                     \dimen2=#2\dimen0
                     \edef #3{\expandafter\c@lean\the\dimen2}}

\def\intdiv #1#2#3{\t@counta=#1
                   \t@countb=#2
                   \ifnum \t@countb<0
                      \t@counta=-\t@counta
                      \t@countb=-\t@countb
                   \fi
                   \t@countd=1                    % record the sign
                   \ifnum \t@counta<0
                      \t@counta=-\t@counta
                      \t@countd=-1
                   \fi
                   \t@countc=\t@counta  \divide \t@countc by \t@countb
                   \t@counte=\t@countc  \multiply \t@counte by \t@countb
                   \advance \t@counta by -\t@counte
                   \t@counte=-1
                   \loop
                     \advance \t@counte by 1
                   \ifnum \t@counte<16                  % loop 16 times
                       \multiply \t@countc by 2           % q=2q
                       \multiply \t@counta by 2           % r=2r
                       \ifnum \t@counta<\t@countb \else   % if ( r >= b )
                         \advance \t@countc by 1          %   q=q+1
                         \advance \t@counta by -\t@countb %   r=r-b
                       \fi
                   \repeat
                   \divide \t@countb by 2         % rounding
                   \ifnum \t@counta<\t@countb     % if ( r >= b/2 ) q=q+1
                     \advance \t@countc by 1
                   \fi
                   \ifnum \t@countd<0             % restore the sign
                     \t@countc=-\t@countc
                   \fi
                   \dimen0=\t@countc sp           % express as a dimension
                   \edef #3{\expandafter\c@lean\the\dimen0}}

\def\coordtopix #1#2{\dimen0=#1\d@dim
                     \dimen2=\d@sc\dimen0
                     \t@counta=\dimen2          % scaled dimension in sp
                     \t@countb=\s@ppix
                     \divide \t@countb by 2
                     \ifnum \t@counta<0         % rounding
                       \advance \t@counta by -\t@countb
                     \else
                       \advance \t@counta by \t@countb
                     \fi
                     \divide \t@counta by \s@ppix
                     #2=\t@counta}

\def\pixtocoord #1#2{\t@counta=#1%
                     \multiply \t@counta by \s@ppix
                     \dimen0=\d@sc\d@dim
                     \t@countb=\dimen0
                     \intdiv \t@counta \t@countb #2}

\def\pixtodim #1#2{\t@countb=#1%
                   \multiply \t@countb by \s@ppix
                   #2=\t@countb sp\relax}

\def\pixtobp #1#2{\dimen0=\p@sfactor pt
                  \t@counta=\dimen0
                  \multiply \t@counta by #1%
                  \ifnum \t@counta < 0             % rounding
                    \advance \t@counta by -32768
                  \else
                    \advance \t@counta by 32768
                  \fi
                  \divide \t@counta by 65536
                  \edef #2{\the\t@counta}}
                  
\newcount\t@counta    \newcount\t@countb   % Use at lowest levels
\newcount\t@countc    \newcount\t@countd
\newcount\t@counte
\newcount\t@pixa      \newcount\t@pixb     % Use for pixel values
\newcount\t@pixc      \newcount\t@pixd

\newdimen\t@xpos      \newdimen\t@ypos

\newcount\xminpix      \newcount\xmaxpix
\newcount\yminpix      \newcount\ymaxpix

\newcount\a@lenpix     \newcount\a@widpix

\newcount\x@pix        \newcount\y@pix
\newcount\x@segoffpix  \newcount\y@segoffpix
\newcount\x@savepix    \newcount\y@savepix

\newcount\s@ppix       % sp/pixel 

\newcount\d@bs

\newcount\t@xdnum
\global\t@xdnum=0

\newbox\t@xdbox

\newwrite\drawfile

\newif\ifm@pending
\newif\ifp@ath
\newif\ifa@st
\newif\ifm@ore
\newif \ift@extonly
\newif\ifp@osinit

\newtoks\everytexdraw

\def\l@paren{(}
\def\a@st{*}

\catcode`\%=12
  \def\p@b {%!}  \def\p@p {%%}
\catcode`\%=14
\catcode`\{=12  \catcode`\}=12  \catcode`\u=1 \catcode`\v=2
  \def\l@br u{v  \def\r@br u}v
\catcode `\{=1  \catcode`\}=2   \catcode`\u=11 \catcode`\v=11

{\catcode`\p=12 \catcode`\t=12
 \gdef\c@lean #1pt{#1}}

\def\sppix#1/#2 {\dimen0=1#2 \s@ppix=\dimen0
                 \t@counta=#1%
                 \divide \t@counta by 2
                 \advance \s@ppix by \t@counta
                 \divide \s@ppix by #1%             % \s@ppix available
                 \t@counta=\s@ppix
                 \multiply \t@counta by 65536       % 1 pt = 65536 sp
                 \advance \t@counta by 32891        % 0.5 bp = 32890.88 sp
                 \divide \t@counta by 65782         % 1 bp = 65781.76 sp
                 \dimen0=\t@counta sp
                 \edef\p@sfactor {\expandafter\c@lean\the\dimen0}}

\def\g@etargxy #1 #2 #3 #4\\#5#6{\def #5{#1}%
                           \ifx #5\empty
                             \g@etargxy #2 #3 #4 \\#5#6%  leading blank
                           \else
                             \def #6{#2}%
                             \def\next {#3}%
                             \ifx \next\empty \else
                               \t@xderror {TeXdraw: invalid coordinate}%
                             \fi
                           \fi}

\def\c@heckast #1{\expandafter
                  \c@heckastll #1\\}
\def\c@heckastll #1#2\\{\def\testit {#1}%
                        \ifx \testit\a@st
                          \a@sttrue
                        \else
                          \a@stfalse
                        \fi}

\def\g@etsympix #1#2{\expandafter
               \ifx \csname #1\endcsname \relax
                 \t@xderror {TeXdraw: undefined symbolic coordinate}%
               \fi
               #2=\csname #1\endcsname}

\def\s@etcsn #1#2{\expandafter
                  \xdef\csname#1\endcsname {#2}}

\def\g@etitem #1 #2\\#3#4{\edef #4{#2}\edef #3{#1}}
\def\a@pppix #1#2{\edef\next {#1}%
                  \ifx \next\empty \else
                    \coordtopix {#1}\t@pixa
                    \ifx #2\empty
                      \edef #2{\the\t@pixa}%
                    \else
                      \edef #2{#2 \the\t@pixa}%
                    \fi
                  \fi}

\def\s@etpospix #1#2{\coordtopix {#1}\x@pix
                     \advance \x@pix by \x@segoffpix
                     \coordtopix {#2}\y@pix
                     \advance \y@pix by \y@segoffpix
                     \u@pdateminmax \x@pix \y@pix}

\def\r@elpospix #1#2{\coordtopix {#1}\t@pixa
                     \advance \x@pix by \t@pixa
                     \coordtopix {#2}\t@pixa
                     \advance \y@pix by \t@pixa
                     \u@pdateminmax \x@pix \y@pix}

\def\r@elupd #1#2{\t@counta=\x@pix
                  \advance\t@counta by #1%
                  \t@countb=\y@pix
                  \advance\t@countb by #2%
                  \u@pdateminmax \t@counta \t@countb}

\def\u@pdateminmax #1#2{\ifnum #1>\xmaxpix
                          \global\xmaxpix=#1%
                        \fi
                        \ifnum #1<\xminpix
                          \global\xminpix=#1%
                        \fi
                        \ifnum #2>\ymaxpix
                          \global\ymaxpix=#2%
                        \fi
                        \ifnum #2<\yminpix
                          \global\yminpix=#2%
                        \fi}

\def\s@avemove #1#2{\x@savepix=#1\y@savepix=#2%
                    \m@pendingtrue
                    \ifp@osinit \else
                      \global\p@osinittrue
                      \global\xminpix=\x@savepix \global\yminpix=\y@savepix
                      \global\xmaxpix=\x@savepix \global\ymaxpix=\y@savepix
                    \fi}

\def\f@lushmove {\global\p@osinittrue
                 \ifm@pending
                   \writetx {\the\x@savepix\space \the\y@savepix\space mv}%
                   \m@pendingfalse
                   \p@athfalse
                 \fi}

\def\f@lushbs {\loop
                 \ifnum \d@bs>0
                   \writetx {bs}%
                   \global\advance \d@bs by -1
               \repeat}
               
\def\h@move #1#2 #3)#4{\move (#2 #3)%
                       \h@text {#4}}
\def\h@text #1{\pixtodim \x@pix \t@xpos
               \pixtodim \y@pix \t@ypos
               \vbox to 0pt{\normalbaselines
                            \t@stuff
                            \kern -\t@ypos
                            \hbox to 0pt{\l@stuff
                                         \kern \t@xpos
                                         \hbox {#1}%
                                         \kern -\t@xpos
                                         \r@stuff}%
                            \kern \t@ypos
                            \b@stuff\relax}}

\def\r@move td:#1 #2#3 #4)#5{\move (#3 #4)%
                             \r@text td:#1 {#5}}
\def\r@text td:#1 #2{\vbox to 0pt{\pixtodim \x@pix \t@xpos
                                  \pixtodim \y@pix \t@ypos
                                  \kern -\t@ypos
                                  \hbox to 0pt{\kern \t@xpos
                                               \rottxt {#1}{\z@sb {#2}}%
                                               \hss}%
                                  \vss}}
\def\z@sb #1{\vbox to 0pt{\normalbaselines
                          \t@stuff
                          \hbox to 0pt{\l@stuff \hbox {#1}\r@stuff}%
                          \b@stuff}}

\ifx \rotatebox\@undefined
  \def\rottxt #1#2{\bgroup
                     #2%
                   \egroup}
\else
  \let\rottxt=\rotatebox
\fi

\ifx \t@xderror\@undefined
  \let\t@xderror=\errmessage
\fi

\def\t@exdrawdef {\sppix 300/in            % 300 pixels/inch
                  \drawdim in              % drawing units are inches
                  \edef\u@nitsc {1}%       % unit scale 1 (has to be set
                  \setsegscale 1           % segment scale 1
                  \arrowheadsize l:0.16 w:0.08
                  \arrowheadtype t:T
                  \textref h:L v:B }

\ifx \includegraphics\@undefined
  \def\t@xdinclude [#1,#2][#3,#4]#5{%
    \begingroup                           % keep definitions local
      \message {<#5>}%
      \leavevmode
      \t@counta=-#1%                      % integer bounding box coordinates
      \t@countb=-#2%
      \setbox0=\hbox{%
        \includegraphics{#5}}%
      \t@ypos=#4 bp%
        \advance \t@ypos by -#2 bp%
      \t@xpos=#3 bp%
        \advance \t@xpos by -#1 bp%
      \dp0=0pt \ht0=\t@ypos  \wd0=\t@xpos
      \box0%
    \endgroup}
\else
  \let\t@xdinclude=\includegraphics
\fi

\def\writeps #1{\f@lushbs
                \f@lushmove
                \p@athtrue
                \writetx {#1}}
\def\writetx #1{\ift@extonly
                  \global\t@extonlyfalse
                  \t@xdpsfn \p@sfile
                  \t@dropen \p@sfile
                \fi
                \w@rps {#1}}
\def\w@rps #1{\immediate\write\drawfile {#1}}

\def\t@xdpsfn #1{%
  \global\advance \t@xdnum by 1
  \ifnum \t@xdnum<10
    \xdef #1{\jobname.ps\the\t@xdnum}
  \else
    \xdef #1{\jobname.p\the\t@xdnum}
  \fi
}
\def\t@dropen #1{%
  \immediate\openout\drawfile=#1%
  \w@rps {\p@b PS-Adobe-3.0 EPSF-3.0}%
  \w@rps {\p@p BoundingBox: (atend)}%
  \w@rps {\p@p Title: TeXdraw drawing: #1}%
  \w@rps {\p@p Pages: 1}%
  \w@rps {\p@p Creator: \TeXdrawId}%
  \w@rps {\p@p CreationDate: \the\year/\the\month/\the\day}%
  \w@rps {50 dict begin}%
  \w@rps {/mv {stroke moveto} def}%
  \w@rps {/lv {lineto} def}%
  \w@rps {/st {currentpoint stroke moveto} def}%
  \w@rps {/sl {st setlinewidth} def}%
  \w@rps {/sd {st 0 setdash} def}%
  \w@rps {/sg {st setgray} def}%
  \w@rps {/bs {gsave} def /es {stroke grestore} def}%
  \w@rps {/fl \l@br gsave setgray fill grestore}%
  \w@rps    { currentpoint newpath moveto\r@br\space def}%
  \w@rps {/fp {gsave setgray fill grestore st} def}%
  \w@rps {/cv {curveto} def}%
  \w@rps {/cr \l@br gsave currentpoint newpath 3 -1 roll 0 360 arc}%
  \w@rps    { stroke grestore\r@br\space def}%
  \w@rps {/fc \l@br gsave setgray currentpoint newpath}%
  \w@rps    { 3 -1 roll 0 360 arc fill grestore\r@br\space def}%
  \w@rps {/ar {gsave currentpoint newpath 5 2 roll arc stroke grestore} def}%
  \w@rps {/el \l@br gsave /svm matrix currentmatrix def}%
  \w@rps    { currentpoint translate scale newpath 0 0 1 0 360 arc}%
  \w@rps    { svm setmatrix stroke grestore\r@br\space def}%
  \w@rps {/fe \l@br gsave setgray currentpoint translate scale newpath}%
  \w@rps    { 0 0 1 0 360 arc fill grestore\r@br\space def}%
  \w@rps {/av \l@br /hhwid exch 2 div def /hlen exch def}%
  \w@rps    { /ah exch def /tipy exch def /tipx exch def}%
  \w@rps    { currentpoint /taily exch def /tailx exch def}%
  \w@rps    { /dx tipx tailx sub def /dy tipy taily sub def}%
  \w@rps    { /alen dx dx mul dy dy mul add sqrt def}%
  \w@rps    { /blen alen hlen sub def}%
  \w@rps    { gsave tailx taily translate dy dx atan rotate}%
  \w@rps    { (V) ah ne {blen 0 gt {blen 0 lineto} if} {alen 0 lineto} ifelse}%
  \w@rps    { stroke blen hhwid neg moveto alen 0 lineto blen hhwid lineto}%
  \w@rps    { (T) ah eq {closepath} if}%
  \w@rps    { (W) ah eq {gsave 1 setgray fill grestore closepath} if}%
  \w@rps    { (F) ah eq {fill} {stroke} ifelse}%
  \w@rps    { grestore tipx tipy moveto\r@br\space def}%
  \w@rps {\p@sfactor\space \p@sfactor\space scale}%
  \w@rps {1 setlinecap 1 setlinejoin}%
  \w@rps {3 setlinewidth [] 0 setdash}%
  \w@rps {0 0 moveto}%
}

\def\t@drclose {%
  \bgroup
    \w@rps {stroke end showpage}%
    \w@rps {\p@p Trailer:}%
    \pixtobp \xminpix \l@lxbp  \pixtobp \yminpix \l@lybp
    \pixtobp \xmaxpix \u@rxbp  \pixtobp \ymaxpix \u@rybp
    \w@rps {\p@p BoundingBox: \l@lxbp\space \l@lybp\space
                              \u@rxbp\space \u@rybp}%
    \w@rps {\p@p EOF}%
  \egroup
  \immediate\closeout\drawfile
}

\catcode`\@=\catamp

\newenvironment{texdraw}{\leavevmode\btexdraw}{\etexdraw}
\def\real{\rm{\bf R}}
\def\uno{\rm{\bf 1}}

\begin{document}
\bibliographystyle{unsrt}

\title{Possible ground states of D-wave condensates in isotropic space}

\author{Yu.M. Gufan$^1$, Al.V. Popov$^1$, G. Sartori$^{2,4}$, V. Talamini$^{3,4}$,
G. Valente$^{2,4}$ and E.B. Vinberg$^5$\\
\small
$^{1}$Rostov State University, Rostov on Don, Russia\\
\small$^2$Dipartimento di Fisica, Universit\`a di Padova, Italy\\
\small $^{3}$Dipartimento di Ingegneria Civile, Universit\`a di
Udine, Italy\\ $^{4}$INFN, Sezione di Padova, Italy\\ \small
$^{5}$ Moscow State University, Moscow, Russia}

%%\small I--35131 Padova, Italy \\ \small (e-mail:
%%gfsartori@padova.infn.it, talamini@padova.infn.it,
%%valente@padova.infn.it)}

%%\date{07/25/2000}

\maketitle

\begin{abstract}
A complete and rigorous determination of the possible ground
states for D-wave pairing Bose condensates is presented. Using an
orbit space approach to the problem, we find 15 allowed phases
(besides the unbroken one), with different symmetries, that we
thoroughly determine, specifying the group-subgroup relations
between bordering phases.
\end{abstract}

%\narrowtext

\section{Introduction}
Superfluidity and superconductivity are justified on the basis of
the macroscopic condensation of Bose quasi-particles. The
classical Bardeen, Cooper and Schrieffer theory for
superconductivity dates 1957. Soon after, a BCS-type transition
was proposed for the Fermi system $^{3}\mbox{\rm He}$ by Anderson
and Morel \cite{1}. Cooper pair formation was thought to occur in
an $L \neq 0$ state, to take into account the hard core nature of
$^{3}\mbox{\rm He}$ atoms interaction. The superfluid phases were
actually observed \cite{1a}, and the nature of p-wave pairing is
now well established for $^{3}\mbox{\rm He}$. The theory of $L
\neq 0$-superfluids is also relevant for ``unconventional''
superconductivity. The high temperature superconducting (HTS) oxides are
anomalous in their non-Fermi liquid normal state properties and
share with heavy fermion superconductors unconventional d-wave
pairing \cite{2,8}. The underlying microscopic mechanism inducing
superconductivity in these materials is still unclear and is one
of today's major challenges \cite{10a,10b}.
%%%%%%%%%%%%%%%%%%%%%%%%%%%%%%%%%%%%%%%%%%%%%%%%%%%%

Such a situation has motivated the efforts at studying the
macroscopic properties of unconventional superconductors through
the Landau theory of phase transitions \cite{GufPR12}. Moreover,
the identification of the order parameter symmetry may be
considered as a preliminary task in the construction of viable
models describing the attractive nature of the pairing
interaction.

Direct experimental evidence of an order parameter unconventional
structure lies on multiple phase transitions. The heavy fermion
compounds U$_{1-x}$Th$_x$Be$_{13}$ display four different
superconducting phases in the $T$-$x$ phase diagram and UPt$_3$
displays three superconducting phases in a $T$-$H$ phase diagram
\cite{2}. Furthermore, from power-law temperature behaviour of
thermodynamic and transport properties (e.g. specific heat,
magnetic penetration depth), a non-trivial node
structure for the gap function may be inferred, compatible with high
$L$-pairing. That is the case also for HTS oxides, for which, at
present, clear proofs are lacking of the existence of more than
one superconducting state.

Actually, it must be pointed out that, unlike
superfluid $^{3}\,\mbox{He}$, which is an isotropic fermion system,
Bloch electrons in a superconductor crystal lattice
exhibit, in general, a reduced finite symmetry; in addition, the existence of
imperfections, even in the cleanest samples, can partly destroy the
gap node structure at low energy. So, the very fact that the
experimental observation cannot at present completely unravel the
node structure of the gap function, reinforces the necessity of classifying
all the possible symmetry breaking schemes.

Moreover, the role of Fermi isotropic space beyond a zero-th
order approximation in phenomenological theories, is still
supported by some recent studies: In HTS oxides, low symmetrical
crystal fields (tetragonal and orthorombic), as well as
spontaneous strain, have weak influence on the temperature of the
superconducting phase transition \cite{11} or on the penetration depth
\cite{12}. This also means that low symmetry crystal fields do not
directly influence condensate states.

The possible ground states of a high $L$ superfluid has been the
object of intense investigations during the 60's and the 70's.
>From the solution of the state equations, Anderson and Morel
\cite{1} and Mermin \cite{14} identified five different phases,
through the minimization of a 4-th
degree Landau potential. Schakel and Bias \cite{16} analyzed the
problem using only group theoretical arguments. Capel and Schakel
\cite{15}, taking advantage of the results in \cite{16}, computed
the possible ground states of condensates driven by p-waves. They
also investigated the consequences of lowering the residual
symmetry resulting from the solutions of this problem by means of
strong spin-orbit interactions. Since, in this case, the Cooper
pair is in a $J=2$ state $(S=1, L=1)$, the order parameter is
represented by a traceless symmetric tensor. The same holds true
for an $(S=0, L=2)$ state order parameter, so those results may be
directly applied to D-wave pairing. According to the authors of
ref. \cite {15}, eleven different phases are allowed. Their
analysis, however, is questionable, since time reversal symmetry
cannot be neglected in the theory of condensate states and the use
of a 4-th degree polynomial Landau free energy strongly limits the
number of phases that would be allowed by the symmetry of the
system.

Our aim in this paper is to give a definitive answer to the
classification of possible symmetry breaking patterns in D-wave
pairing Bose condensate, in the framework of the Landau theory of
phase transitions. To this end we make use of a geometrical
invariant theory approach to the problem, proposed in
ref.~\cite{SA}. The strategy is to exploit a set of basic
invariant polynomials of the symmetry group of the system, as
fundamental variables in the description of the phase space of the
system and in the minimization procedure of the free energy
\cite{23}.

\section{The orbit space approach}

Let us briefly recall some basic elements of the orbit space
approach \cite{NC}. To this end, we shall denote by $x\in {\bf
R}^n$ a vector order parameter, transforming linearly and
orthogonally under the compact real symmetry group $G$, and by
$\Phi(\alpha; x)$ the $G$-invariant free energy, expressed also in
terms of state variables $\alpha$. The points $x_0(\alpha)$, where
the function $\phi_\alpha(x) = \Phi(\alpha; x)$ takes on its
absolute minimum, correspond to the stable phase of the system,
whose symmetry is determined by the isotropy subgroup, $G_{x_0}$,
of $G$ at $x_0$. Owing to $G$-invariance, the stationary points of
the free energy are degenerate along $G$-orbits. Since the
isotropy subgroups of $G$ at points of the same orbit are
conjugate in $G$, only the conjugate class, $[G_{x_0}]$, of
$G_{x_0}$ in $G$, {\em i.e.} the {\em symmetry} (or {\em orbit-type}) of the orbit through
$x_0$, is physically relevant.

The set of all $G$-orbits, endowed with the quotient topology and
differentiable structure, forms the {\em orbit space}, ${\bf
R}^n/G$, of $G$ and the subset of all the $G$-orbits with a given
symmetry forms a {\em stratum} of ${\bf R}^n/G$. Phase transitions
take place when, by varying the values of the $\alpha$'s, the
point $x_0(\alpha)$ is shifted to an orbit lying on a different
stratum. If $\Phi(\alpha;x)$ is a sufficiently general function of
the $\alpha$'s, by varying these parameters, it is possible to
shift $x_0(\alpha)$ on whichever stratum of ${\bf R}^n/G$. So, the
strata are in one-to-one correspondence with the symmetry phases
allowed by the $G$-invariance of the free energy. On the contrary,
extra restrictions on the form of the free energy function, not
coming from G-symmetry requirements (e.g., the assumption that the
free energy is a polynomial of low degree), can limit the number
of allowed phases.

Being constant along each $G$-orbit, the free energy may be
conveniently thought of as a function defined in the orbit space
of $G$. This fact can be formalized using some basic results of
invariant theory. In fact, the G-invariant polynomial functions
separate the $G$-orbits, meaning that, for any two distinct
$G$-orbits, there is at least a polynomial $G$-invariant function
assuming different values on them. Moreover, every G-invariant  polynomials can be built
as real polynomial functions of a finite set, $\{p_1(x), \dots
,p_q(x)\}$, of basic polynomial invariants ({\em integrity basis
of the ring of $G$-invariant polynomials}), which need not, for
general compact groups, be algebraically independent. The number
of algebraically independent elements in a {\ em minimal} set of basic
polynomial invariants is $n - \nu$, where $\nu$ is the dimension
of the generic (principal) orbits of $G$. Information on the
number and degrees of a minimal set of basic invariants, and the
degrees of the algebraic relations ({\em syzygies}) among them,
can be inferred from the M\"olien function of $G$.

Let us call $q_0$ the number of independent elements of the set
$\{p\}$. The range of the {\em orbit map}, $x \mapsto p(x) =
(p_1(x), \dots ,p_q(x))\in {\bf R}^q$, yields a realization of the
orbit space of the linear group $G$, as a connected semi-algebraic
surface, {\em i.e.} a subset of ${\bf R}^q$, determined by algebraic
equations and inequalities. The orbit space of $G$ is, therefore,
a closed and connected region of a $q_0$-dimensional algebraic
surface, delimited by lower dimensional semi-algebraic surfaces.

Like all semi-algebraic sets, the orbit space of $G$ presents a
natural {\em stratification}, since it can be considered as the
disjoint union of semi-algebraic subsets of various dimensions
({\em geometrical strata}), each stratum being in the border of a
higher dimensional stratum, but for the highest dimensional one,
which is unique ({\em principal stratum}). The connected
components of the {\em symmetry strata} are in one-to-one
correspondence with the {\em geometrical strata}. The symmetries
of two bordering strata are related by a group subgroup relation
and the lower dimensional stratum has a larger symmetry.

The orbit space can be identified with the semi-algebraic variety, $S$,
formed by the points $p\in {\bf R}^q$, satisfying the following
conditions i) and ii) \cite{SA,PS}:
\begin{itemize}
\item[i)] $p$ lies on the surface, $Z$, defined by the syzygies;
\item[ii)] the $q\times q$ matrix $\widehat P(p)$,
defined by the relation

\begin{equation} \widehat P_{ab}(p(x)) = \sum_{j=1}^n\partial_j
p_a(x)\,\partial_j p_b(x),\quad \forall x\in {\bf R}^n \end{equation} is
positive semidefinite at $p$.
\end{itemize}

The relations defining the strata can be obtained as positivity
and rank conditions on the matrix $\widehat P(p)$ and the minimum
of $\Phi(\alpha;x)$ can be computed as a constrained minimum in
orbit space of the function $\widehat \Phi(\alpha;p)$,

\begin{equation}
\widehat \Phi(\alpha;p(x))= \Phi(\alpha;x),\;\forall x\in {\bf R}^{n}
\end{equation}
or from the solutions of the equation \cite{SA}

\begin{equation}
\sum_{b=1}^q\widehat P_{ab}(p)\,\partial_b \widehat{\Phi}(\alpha ; p)=0, \qquad a=1,\dots
,q,\label{100}
\end{equation}
which is equivalent to the state equation, $\partial \Phi(\alpha;x)/\partial x_j = 0$, $j=1,\dots ,n$.

\section{Symmetry of the allowed D-wave condensate states in
isotropic space}

The formation of D-wave condensate states breaks the symmetry of
the isotropic 3-dimensional space, which corresponds to the group
{\bf O}$_3\otimes${\bf U}$_1\times \left\langle {\cal
T}\right\rangle$, where {\bf O}$_3$ is the complete rotation
group, {\bf U}$_1$ is the group of gauge transformations and
$\left\langle {\cal T}\right\rangle $ is the group generated by
the time reversal operator ${\cal T}$.

The symmetry of the allowed D-wave condensate ground states is
defined by the relative values of the complex coefficients in the
decomposition of the gap-function, $\Delta$, in terms of spherical
harmonics with $L = 2$:

\begin{equation} \Delta(\theta,\phi) = \sum_{m=-2}^2 D_m\,Y_2^m(\theta,\phi)
\end{equation}

The set of functions $\{Y_2^m,Y_2^{m\,*}\}$ yields a basis of a
ten-dimensional (10\,D) space hosting a real representation of the
symmetry group {\bf O}$_3\otimes${\bf U}$_1\times \left\langle {\cal T}\right\rangle$.
A general element, $\gamma$, of the group will be denoted
by a triple $\gamma =(\rho,\,e^{i\phi},\,\epsilon)$, where, $\rho \in
{\rm\bf O}_3$, $0\le\phi <2\pi$ and $\epsilon =- 1$, or $+1$
according as a time reflection is involved in the transformation,
or not.

The action of $G$ can be transferred to a real irreducible action
on the 10\,D vector formed by the coefficients $\{D_{2},\dots ,D_{-2},
D_{-2}^*,\dots ,D_2^*\}$. The representation of $G$ thus obtained can
be realized in the 10\,D real vector space of a couple of two
independent, real, second rank, symmetric, traceless tensors,
$X^{(1)}_{ij}$ and $X^{(2)}_{ij}$, $i,j=1,2,3$, which can be
considered as the real and imaginary parts of a complex $3\times
3$ matrix $\psi$, whose elements will be written in terms of five complex coordinates, $z_j$:

\begin{equation}
z_j = x_j + i\, x_{5+j},\qquad j=1,\dots ,5; \qquad x_i\in{\bf R},
\end{equation}

\begin{equation}
\psi=\frac 1{\sqrt{2}}\left(
\begin{array}{ccc}
z_2 + \frac{\displaystyle z_5}{\sqrt 3} & z_1                                      & z_3  \\
z_1                                     & -z_2 + \frac{\displaystyle z_5}{\sqrt{3}}& z_2  \\
z_3                                     &  z_2                                     &  -\frac{\displaystyle 2\,z_5}{\sqrt{3}}
\end{array} \right).
\end{equation}
The coordinates $D_\alpha$ are connected to the $z_j$ by the following relations:

\begin{equation}
D_2    = {\displaystyle -\frac{i\,z_1 + z_2}{\sqrt 2}},\quad D_1 = {\displaystyle \frac{i\,z_3 + z_4}{\sqrt
2}},\quad D_0 = z_5,\quad D_{-1} = {\displaystyle \frac{i\,z_3 - z_4}{\sqrt
2}},\quad D_{-2} = {\displaystyle \frac{i\,z_1 - z_2}{\sqrt 2}}\,.
\end{equation}

The matrix $\psi$ transforms in the following way under a general transformation $\gamma=(\rho,\phi,\epsilon)\in G$:

\begin{equation}
 \gamma\cdot\psi = e^{i\phi}\,\rho\,\psi' \,\rho^{\rm T},\qquad \gamma \in G,
 \label{trasf}
 \end{equation}
where $\psi'= \psi$ or $\psi^*$, according as $\epsilon =+1$ or $-1$ and the apex T denotes transposition.
As a consequence, the group $G$ acts as a group of
linear, real, {\em orthogonal} transformations on the vector order parameter
$x\in {\bf R}^{10}$.

The kernel of the representation of $G$ just defined is the group generated by the space reflection. So, it will
not be restrictive to assume that the symmetry group is $G
={\rm\bf SO}_3\otimes{\rm\bf U}_1\times \left\langle{\cal T}\right\rangle$
and, when speaking of $G$, in the following, we shall always refer to this linear group acting in the vector space
$\real^{10}$.

The linear group $G$ has a trivial principal isotropy subgroup
(the isotropy subgroup of generic points of ${\bf R}^{10}$), thus
the principal $G$-orbits have the same dimensions as $G$ and its {\em orbit space}, {\em i.e.}, the
quotient space ${\bf R}^{10}/G$, has dimensions $q_0 = 10 - 4 = 6$.

The M\"olien function  of $G$, $M(\eta)$, can be calculated in the
form of an invariant Haar integral over $G$ (see, for instance,
\cite{33,3334}):

\begin{equation}
M(\eta ) = \int_G\frac {d\mu(g)}{{\rm det}\left(\uno -
\eta\,g\right)},\qquad |\eta| < 1, \label{8}
\end{equation}
where $\mu(g)$ is a normalized invariant measure on the group $G$,
the integration is over the whole group $G$ and $g\in G$. An explicit
calculation of the integral leads to

\begin{equation}
M(\eta ) = \frac{\eta ^{20} + \eta ^{12} + \eta ^{10} + \eta ^8 +
1}{(1 - \eta ^2)(1 - \eta ^4)^2(1 - \eta ^6)^2(1 - \eta
^8)}.\label{10}
\end{equation}
Equation (\ref{10}) yields the following indications, whose validity has been checked through
direct calculations:

\begin{enumerate}
\item A minimal integrity basis for the linear
group $G$ contains nine elements, $\{p_1,\dots , p_9\}$ with
degrees $(d_1,\dots ,d_9)=(2, 4, 4, 6, 6, 8, 8, 10, 12)$.

\item The invariants $p_i$ are connected by five independent syzygies of degrees 16, 18, 20, 22
and 24.

\item The most general $G$-invariant polynomial,  like a general non-equilibrium polynomial Landau potential,
$\widehat\Phi(\alpha;p)$, can be written as a polynomial function
of the elements of the integrity basis $\{p_i\}_{i=1 \ldots 9}$,
in terms of five arbitrary polynomials \cite{Stanley}, $Q_i=Q_i(\alpha;p_1,\dots ,p_6)$, $i=0,\dots ,4$:

\begin{equation}
\widehat\Phi = Q_0 + Q_1\, p_7 + Q_2\, p_8 + Q_3\, p_9 + Q_4\, p_7\,p_9.\end{equation}
\end{enumerate}

The elements of the minimal integrity basis can be chosen in the following
form:

\begin{equation}
\begin{array}{ll}
\begin{array}{rcl}
p_1&=&{\rm Tr}(\psi\psi^*)=\sum_{i=1}^{10}x_i^2,\\
p_2&=& {\rm Tr}\left[(\psi\psi^*)^2\right],\\ p_3&=& |{\rm Tr}(\psi^2)|^2, \\
p_4 &=&|{\rm Tr}(\psi^3)|^2,\\
p_5 &=& |{\rm Tr}(\psi^2\psi^*)|^2,
\end{array}
&
\begin{array}{rcl}
p_6 &=& \Re\left[{\rm Tr}(\psi^2){\rm Tr}(\psi^2\psi^*){\rm Tr}(\psi^{*3})\right],\\
p_7 &=& \Re\left[{\rm Tr}(\psi^{*2})\left({\rm Tr}(\psi^2\psi^*)\right)^2\right],\\
p_8 & =& \Re\left[{\rm Tr}(\psi^2\psi^*){\rm Tr}(\psi^3)\left({\rm Tr}(\psi^{*2})\right)^2\right],\\
p_9 &=& \Re\left[\left({\rm Tr}(\psi^2)\right)^3\,\left({\rm Tr} (\psi^{*3})\right)^2\right.]\,.
\end{array}
\end{array}
\end{equation}

Using these definitions, we have determined the explicit form of
the syzygies, of the $\widehat P$--matrix elements, of the
equations and inequalities determining the strata in the orbit
space of $G$. For each stratum, denoted by $S^{(d,r)}$, where $d$ denotes the dimension and $r$ is an enumeration
index, we have picked up a ``typical point" and determined the corresponding isotropy subgroup of $G$.
All these results are essential pre-requisites for a rigorous
calculation of the minima of a $G$-invariant polynomial along the
lines indicated in the Introduction.

Part of our results are resumed in Tables~\ref{T1}, \ref{T2}
and \ref{T3} and in Fig. 1, where the group--subgroup relations among the
symmetries of bordering strata are also specified.

The explicit expressions of the syzygies and of the elements of
the $\widehat P(p)$ matrix  and the relations defining higher
dimensional strata would require too much space to be written down
here.

As a simple example of the effectiveness of our approach, we have also repeated Mermin's calculation
of the minimum of a general fourth degree polynomial $G$-invariant free energy

\begin{equation}
\widehat\Phi^{(4)}(p)=\alpha_0\,\frac{p_1^2}2 + \sum_{j=1}^3 \alpha_j \,p_j,\label{11}
\end{equation}
in the additional assumptions that it is bounded below and has a local maximum at the origin ($\alpha_1<0$).
The $\alpha_i$'s are connected to the parameters $\alpha, \beta_1,\beta_2,\beta_3$ used by Mermin by the following relations:

\begin{equation}
\alpha_0=2\,\beta_2+\beta_3,\qquad  \alpha_1= \alpha,\qquad \alpha_2
= -\frac{\beta_2}2,\qquad \alpha_3 = \beta_1 +
\frac{\beta_3}4.\label{mermin}
\end{equation}

With the definitions:

\begin{equation}
\tilde p_i = {\displaystyle \frac {p_i}{p_1^{d_i/2}}},\qquad \tilde p = (\tilde p_2,\dots ,\tilde
p_9),\label{ptilde}
\end{equation}
and

\begin{equation}
\Delta(\tilde p) = \alpha_0 + 2\,\alpha_2\,\tilde p_2 + 2\,\alpha_3\,\tilde
p_3,\label{delta}
\end{equation}
the polynomial $\widehat\Phi^{(4)}(p)$ can be put in the following convenient form:

\begin{equation}
\widehat\Phi^{(4)}(p) = \frac {p_1^2}2\,\Delta(\tilde p) + \alpha_1\, p_1.\label{Delta}
\end{equation}

Since, owing to its definition, $p_1$ ranges over the whole non negative real numbers, $\widehat\Phi^{(4)}(p)\big|_{p\in
 S}$ is bounded below ({\em stability condition}) if and only if the minimum, $\delta$, of $\Delta(\tilde p)
\big|_{p_1=1}$ is positive. Being the minimum of the r.h.s. of (\ref{Delta}), thought of as a function only of
$p_1 \ge 0$,  equal to $-\alpha_1/(2\Delta)$, the absolute minimum of $\widehat\Phi^{(4)}(p)\big|_{p\in
S}$ is $-\alpha_1^2/(2\delta)$. In this way, the problems of rendering explicit the stability condition and evaluating
the minimum of $\widehat \Phi(p)\big|_{p\in S}$ are reduced to the calculation of $\delta$.

The absolute minimum of $\delta$ in each singular stratum can be easily obtained from the equations of the strata, listed in
Tables \ref{T1} and \ref{T2} for strata with dimensions $< 3$. For the principal stratum, it is easier to solve the
projection of equation (\ref{100}) in the unit sphere ($\sum_{i=1}^{10}x_i^2 = p_1 = 1$), which, using (\ref{ptilde}),
can be written in the form \cite{ST}

\begin{equation}
\sum_{j=2}^9 P_{ij}(\tilde p) \,\frac{\partial \Delta(\tilde p)}{\partial \tilde p_j} = 0
\end{equation}
and to select, subsequently, the solutions lying in the principal stratum.

A comparison of the values of the minima in the different strata, obtained in this way, leads to the
results resumed in Table~\ref{T4} and illustrated in Figure 2. Owing to the low degree of the polynomial defining
$\widehat\Phi^{(4)}(p)$ in (\ref{11}), and the consequent low number of free parameters $\alpha$, the absolute minimum
presents strong degeneracy, particularly for special values of the $\alpha$'s.
If these special values are excluded, spontaneous breaking of the symmetry can generate only five distinct phases out of
fifteen permitted by the $G$-symmetry; some of them are unstable. For no non trivial values of $(\alpha_2,\alpha_3)$ does
the absolute minimum lie on the stratum $S^{(2,2)}$.

For general values of the $\alpha$'s, our results are in agreement with Mermin's ones. Let us add a few words
about the perturbative stability of the three degenerate phases in the region ${\cal R}_2$ (see Table~\ref{T4}).
For $(\alpha_0,\alpha_2,\alpha_3)\in{\cal R}_2$, the addition to the free energy, $\Phi^{(4)}(p)$, of a ''small"
perturbation, consisting in an invariant polynomial of degree six, $\Theta^{(6)}=\alpha_4\,p_4 + \alpha_5\,p_5$,
splits the three degenerate minima
determined by the 4-th degreee term\footnote{The inclusion of terms of 6-th degree, depending only on
$p_1,p_2$ and $p_3$ would be useless for splitting the degenerate minima, so these terms will be neglected, with no loss
of generality.}. This is easy to check, at least in the additional assumption that the
perturbation leaves the absolute minimum in one of the strata corresponding to the degenerate phases. In fact,
at the first perturbative order, one obtains from Tables~\ref{T1} and \ref{T2} the following shifts,
$\Theta^{(6)}_{(d,r)}$, in the values of the 6-th order free energy at the points where $\Phi^{(4)}(p)$ takes on its
degenerate absolute minimum under consideration:

\begin{equation}
\Theta^{(6)}_{(1,2)} = \left(\frac{\alpha_1}{\delta}\right)^3\,\frac{\alpha_4 + \alpha_5}6,\qquad
\Theta^{(6)}_{(1,3)} = 0,\qquad \Theta^{(6)}_{(2,4)} =  \left(\frac{\alpha_1}{\delta}\right)^3\,
\left(\alpha_4 + \alpha_5\right)\,\xi,
\end{equation}
where $0< \xi < 1/6$.

Since $-\alpha_1/\delta > 0$, the absolute minimum will be perturbatively stable on $S^{(1,2)}$ or,
respectively, $S^{(1,3)}$, according as $\alpha_4+\alpha_5$ is negative or positive.

As stressed in the Introduction, the difficulties mentioned above can be overcome if one puts less restrictive
upper limits to the degree of the polynomial describing the free energy. It is trivial, for instance, to realize
that the following class of bounded below polynomial functions \cite{NC} have a vanishing maximum at the origin of
$\real^{10}$
and display an absolute minimum at the arbitrarily chosen point $\bar p\in S$:

\begin{equation}
\sum_{i=1}^9\,\alpha_i\left[\left(p_i - \bar p_i\right)^{2n_i} - \bar p_i^{2n_i}\right],
\end{equation}
where the $\alpha$'s are positive constants and the $n$'s are positive integers.

The physical implications of our results and the derivation of a more realistic form of the free energy will
be discussed in forthcoming papers.

\acknowledgments This paper is partially supported by RFBR, INFN
and MURST.

%%%%%%%%%%%%%%%%%%%%%%%%%% TABLES %%%%%%%%%%%%%%%%%%%%%%%%%%%%%%%%%%%%%%%%%%%%

\begin{table}
 \caption{Relations defining strata, $S^{(1,r)}$, of dimensions 1 in orbit space. For $2\leq i \leq 9$,
 $q_i=p_i/(p_1^{d_i/2})$ and $d_i$
 denotes the degree of the polynomial $p_i(x)$ ($x$ is the order parameter).
 \label{T1}}

\begin{tabular}{ccccccccc}
$r$ & $\tilde p_2$ & $\tilde p_3$ & $\tilde p_4$ & $\tilde p_5$ & $\tilde p_6$ & $\tilde p_7$ & $\tilde p_8$ & $\tilde p_9$ \\ \hline
1   &  1           &    0         &     0        &   0          &   0          &   0          &   0          &      0       \\
2   & $1/2$        &    1         &   $1/6$      & $1/6$        & $1/6$        & $1/6$        & $1/6$        &    $1/6$     \\
3   & $1/2$        &    1         &     0        &   0          &   0          &   0          &   0          &    0         \\
4   & $1/3$        &    0         &   $1/3$      &   0          &   0          &   0          &   0          &    0         \\
5   & $1/2$        &    0         &     0        &   0          &   0          &   0          &   0          &    0         \\
\end{tabular}
\end{table}

\begin{table}
 \caption{Relations defining strata, $S^{(2,r)}$, of dimensions
 $2$ in orbit space. The $\tilde p_i$'s are defined as
 in Table~\ref{T1} and $\epsilon = \pm 1$. \label{T2}}

\begin{tabular}{cccccc}
$\tilde p\backslash r$ & $1$                                & 2
& $3$                                       & 4             & 5
\\ \hline $\tilde p_2$           & $(2+\xi^2)/6$
& $1/2$                        & $(2+\xi^2)/6$
& $1/2$         & $\xi$          \\ $\tilde p_3$           & 0
& $\xi$                        & $\xi^2$
&  1            & $2-2\,\xi$     \\ $\tilde p_4$           &
$(2-\xi)^2(1+\xi)/12$              & 0
& $(2-\xi)^2(1+\xi)/12$                     & $\xi$         & 0
\\ $\tilde p_5$           & 0                                  & 0
& $\xi^2(1+\xi)/12$                         & $\xi$         & 0
\\ $\tilde p_6$           & 0                                  & 0
& $(2-\xi)(1+\xi)\,\xi^2/12$                & $\xi$         & 0
\\ $\tilde p_7$           & 0                                  & 0
& $(1+\xi)\,\xi^3/12$                       & $\xi$         & 0
\\ $\tilde p_8$           & 0                                  & 0
& $(2-\xi)(1+\xi)\,\xi^3/12$                & $\xi$         & 0
\\ $\tilde p_9$           & 0                                  & 0
& $(2-\xi)^2(1+\xi)\,\xi^3/12$              & $\xi$         & 0
\\ $\xi$ range            & $]0,\frac{1+3\,\epsilon}2[$        &
]0,1[                        & $]0,\epsilon [$ & $]0,\frac 16[$&
$]\frac 12,1[$ \\
\end{tabular}
\end{table}

\begin{table}
 \caption{Possible symmetry strata, $S^{(d,r)}$, for
 D-wave driven pairing in isotropic space ($d$ denotes the dimension of the stratum and $r$ is
 a secondary enumeration index).
 From left to right, the columns refer to phase reference numbers according to
 our($(d,r)$) and, respectively, ref.[12] (${\cal N}$)
 classification, order ($|H|$) and residual symmetry group ($H$), complex coordinates
 ($z_j=x_j+i\,x_{5+j}$,
 $j=1,\dots ,5$) of the $H$-invariant superconducting vector order parameter and corresponding values of the partial wave
 amplitudes $\scriptstyle (D_2,\dots,D_{-2})$. In columns five and six, the $t$'s are real coordinates, while the $v$'s are
 complex ones.
 The notations are the same as in [19], $R_z(\phi)$ denotes
a clockwise proper rotation of the vectors, about the $z$-axis and {\bf O}$_2^z=\{{\rm R}_z(\phi)\}_{0\le\phi <2\pi}\cup
\{C_{2x} {\rm R}_z(\phi)\}_{0\le\phi <2\pi}$. \label{T3}}
%%%%%%%%%%%
\begin{tabular}{cccccc}
$             (d,r)$ & ${\cal N}      $     & $ |H|$ & $H$ &
$\small       (z_1,\dots ,z_5            )$           & $\small
\sqrt 2\left(D_2,D_1,D_0,D_{-1},D_{-2}\right)$
\\ \hline $\scriptstyle (1,1)$ & $\scriptstyle II  $  &
$\scriptstyle \infty$    & $\scriptstyle \left\langle C_{2x}{\cal
T}\right\rangle \times\{{\rm R}_z(\phi)\,{\rm U}_1(2\phi) \}_\phi$
& $\scriptstyle (it,-t,0,0,0               )$           &
$\scriptstyle (2 t,0,0,0,0 )$ \\ $\scriptstyle (1,2)$ &
$\scriptstyle VIII$  & $\scriptstyle \infty$    & $\scriptstyle
{\rm\bf O}_2^z\otimes\left\langle {\cal T}\right\rangle$
& $\scriptstyle (0,0,0,0,t                 )$           &
$\scriptstyle (0,0,t\sqrt 2,0,0 )$ \\ $\scriptstyle (1,3)$ &
$\scriptstyle IX,X$  & $\scriptstyle 16$        & $\scriptstyle
\left\langle C_{2x},{\cal T},C_{4z}{\rm U}_1(\pi)\right\rangle$
& $\scriptstyle (0,t,0,0,0                 )$           &
$\scriptstyle (-t,0,0,0,-t )$ \\ $\scriptstyle (1,4)$ &
$\scriptstyle XI  $  & $\scriptstyle 24$        & $\scriptstyle
\left\langle C_{2x},C_{2a}{\cal T},C_{3\delta}\,{\rm U}_1(4
\pi/3)\right\rangle $           & $\scriptstyle (0,-it,0,0,t
)$           & $\scriptstyle (it,0,t\sqrt 2,0,it )$ \\
$\scriptstyle (1,5)$ & $\scriptstyle I   $  & $\scriptstyle
\infty$    & $\scriptstyle \left\langle C_{2x}{\cal
T}\right\rangle\times\{{\rm R}_z(\phi)\,{\rm U}_1(- \phi) \}_\phi$
& $\scriptstyle (0,0,t,-it,0               )$           &
$\scriptstyle (0,0,0,2\,it,0 )$ \\ $\scriptstyle (2,1)$ &
$\scriptstyle V   $  & $\scriptstyle 6$         & $\scriptstyle
\left\langle C_{2x}{\cal T}, C_{3z}{\rm U}_1(4\pi/3)\right\rangle$
& $\scriptstyle (it_1,-t_1,t_2,-it_2,0     )$           &
$\scriptstyle (t_1,0,0,i\,t_2,0 )$ \\ $\scriptstyle (2,2)$ &
$\scriptstyle IV  $  & $\scriptstyle 4$         & $\scriptstyle
\left\langle C_{2x}{\cal T}, C_{2z}{\rm U}_1(\pi)\right\rangle$
& $\scriptstyle (0,0,t_1,it_2,0            )$           &
$\scriptstyle (0,it_1+i\,t_2,0,it_1-it_2 )$ \\ $\scriptstyle
(2,3)$ & $\scriptstyle --  $  & $\scriptstyle 8$         &
$\scriptstyle \left\langle C_{2x}, C_{4z}{\cal T}\right\rangle$
& $\scriptstyle (0,it_1,0,0,t_2            )$           &
$\scriptstyle (-it_1,0,t_2\sqrt 2,0,-it_1 )$ \\ $\scriptstyle
(2,4)$ & $\scriptstyle VII $  & $\scriptstyle 8$         &
$\scriptstyle \left\langle C_{2x}, C_{2z}, {\cal T}\right\rangle$
& $\scriptstyle (0,t_1,0,0,t_2             )$           &
$\scriptstyle (-t_1,0,t_2\sqrt 2,0,-t_1 )$ \\ $\scriptstyle (2,5)$
& $\scriptstyle VI  $  & $\scriptstyle 8$         & $\scriptstyle
\left\langle C_{2x}{\cal T}, C_{4z}{\rm U}_1(\pi)\right\rangle$
& $\scriptstyle (it_1,t_2,0,0,0            )$           &
$\scriptstyle (t_1-t_2,0,0,0,-t_1-t_2 )$ \\ $\scriptstyle (3,1)$ &
$\scriptstyle III $  & $\scriptstyle 4$         & $\scriptstyle
\left\langle C_{2z}, C_{2x}{\cal T}\right\rangle$
& $\scriptstyle (it_1,t_2,0,0,t_3          )$           &
$\scriptstyle (t_1-t_2,0,t_3\sqrt 2,0,-t_1-t_2 )$ \\ $\scriptstyle
(3,2)$ & $\scriptstyle --  $  & $\scriptstyle 4$         &
$\scriptstyle \left\langle C_{2x}, C_{2z}\right\rangle$ &
$\scriptstyle (0,v_1,0,0,v_3             )$           &
$\scriptstyle (-v_1,0,v_2\sqrt 2,0,-v_1 )$ \\ $\scriptstyle (4,1)$
& $\scriptstyle --  $  & $\scriptstyle 2$         & $\scriptstyle
\left\langle C_{2x}{\cal T}\right\rangle$ & $\scriptstyle
(it_1,t_2,t_3,it_4,t_5     )$           & $\scriptstyle
(t_1-t_2,it_3+it_4,t_5\sqrt 2,it_3-it_4,-t_1-t_2 )$ \\
$\scriptstyle (4,2)$ & $\scriptstyle --  $  & $\scriptstyle 2$
& $\scriptstyle \left\langle C_{2z}\right\rangle$ & $\scriptstyle
(v_1,v_2,0,0,v_3           )$           & $\scriptstyle
(-iv_1-v_2,0,v_3\sqrt 2,0,iv_1-v_2 )$ \\ $\scriptstyle (6,1)$ &
$\scriptstyle --  $  & $\scriptstyle 1$         & $\scriptstyle
\{{\bf 1} \}$ & $\scriptstyle (v_1,v_2,v_3,v_4,v_5       )$
& $\scriptstyle (-iv_1-v_2,iv_3+v_4,v_5\sqrt 2,iv_3-v_4,iv_1-v_2
)$ \\
\end{tabular}
%%%%
\end{table}

\begin{table}
\caption{Absolute minimum, $\widehat\Phi^{(4)}_{\rm min}=-\alpha_1^2/(2\delta)$, of a general, bounded below, $G$-invariant
4-th degree polynomial, $\widehat\Phi^{(4)}(\alpha,p)=\alpha_0 p_1^2/2 + \sum_{j=1}^3 \alpha_j p_j$, $\alpha_1<0$, and
hosting strata, $S^{(d,r)}$, as functions of the coefficients $\alpha$.
The denomination of the strata is the same as in Table~\ref{T3}. \label{T4} }
\begin{tabular}{ccc}
$\alpha$ range                                                 & $\delta$                           & $(d,r)$   \\ \hline
${\cal R}_1:\ {\rm Max}(0,-3\alpha_0/2,-6\alpha_3)<\alpha_2$                & $\alpha_0 + 2\alpha_2/3$          & (1,4)          \\
${\cal R}_2:\ -6\alpha_3>\alpha_2>{\rm Max}(-\alpha_0-2\alpha_3,2\alpha_3)$ & $\alpha_0 + \alpha_2 + 2\alpha_3$ & (1,2), (1,3), (2,4) \\
${\cal R}_3:\ -\alpha_0/2 < \alpha_2 < {\rm Min}(0,2\alpha_3)$              & $\alpha_0 + 2\alpha_2$            & (1,1)          \\
${\cal R}_{13}:\ 0=\alpha_2<{\rm Min}(\alpha_0,\alpha_3)$                   & $\alpha_0$                        & (1,1), (1,4), (1,5), (2,1), (3,1), (4,1)\\
${\cal R}_{12}:\ {\rm Max}(-3\alpha_0/2,0)<\alpha_2=-6\alpha_3$             & $\alpha_0-4\alpha_3$              & (1,2), (1,3), (1,4), (2,3), (2,4), (3,2)\\
${\cal R}_{23}:\ -\alpha_0/2<\alpha_2=2\alpha_3<0$                          & $\alpha_0+4\alpha_3$              & (1,1), (1,2), (1,3), (2,4), (2,5)\\
${\cal R}_{123}:\ \alpha_2 = \alpha_3 = 0 < \alpha_0$                       & $\alpha_0$                        & all, except (0,1)   \\
\end{tabular}
\end{table}

\begin{center}
\begin{figure}
\caption{Possible phase transitions between bordering strata, connected, in the figure, by continuous sequences of one or
more arrows.
 The notations are the same as in Table~\ref{T3}.}
\label{F1}

\begin{texdraw}

\drawdim mm

\move(80 180) \textref h:C v:C \htext{\ }

\move(80 150) \textref h:C v:C
\htext{\framebox{\scriptsize$\def\arraystretch{0.8}\begin{array}{c}
S^{(0)}=\{0\}\\{\rm\bf SO}_3\otimes{\rm\bf U}_1\times{\cal
T}\end{array}$}}

\move(80 144) \avec(4 127) \move(80 144) \avec(47 127) \move(80
144) \avec(85 127) \move(80 144) \avec(115 127) \move(80 144)
\avec(150 127)

\textref h:C v:C \htext(4
120){\framebox{\scriptsize$\def\arraystretch{0.8}\begin{array}{c}
S^{(1,5)}\\ {[}\left\langle C_{2x}{\cal T}\right\rangle
\times\{{\rm R}_z(\phi){\rm U}_1(-\phi)\}_\phi {]} \end{array}$}}

\textref h:C v:C \htext(47
120){\framebox{\scriptsize$\def\arraystretch{0.8}\begin{array}{c}
S^{(1,4)}\\ {[}\left\langle C_{2x},C_{3\delta}{\rm
U}_1(\frac{4\pi}{3}),C_{2a}{\cal T}\right\rangle {]}
\end{array}$}}

\textref h:C v:C \htext(85
120){\framebox{\scriptsize$\def\arraystretch{0.8}\begin{array}{c}
S^{(1,3)}\\ {[}\left\langle C_{2x},C_{4z}\,{\rm U}_1(\pi),{\cal
T}\right\rangle {]} \end{array}$}}

\textref h:C v:C \htext(115
120){\framebox{\scriptsize$\def\arraystretch{0.8}\begin{array}{c}
S^{(1,2)}\\ {[}{\rm\bf O}_2^z\times\left\langle {\cal
T}\right\rangle {]} \end{array}$}}

\textref h:C v:C \htext(150
120){\framebox{\scriptsize$\def\arraystretch{0.8}\begin{array}{c}
S^{(1,1)}\\
 {[}\left\langle C_{2x}{\cal T}\right\rangle  \times  \{{\rm R}_z(\phi){\rm
 U}_1(2\phi)\}_\phi
%_{0\le \phi <2\pi}
{]} \end{array}$}}

\move(4 114) \avec(10 97)\move(4 114)  \avec(47 97) \move(47 114)
\avec(10 97)\move(47 114)  \avec(79 97) \move(85 114) \avec(47 97)
\move(85 114)  \avec(79 97) \move(85 114) \avec(111 97)\move(85
114)  \avec(146 97) \move(115 114) \avec(79 97)\move(115 114)
\avec(111 97) \move(150 114) \avec(10 97)\move(150 114)  \avec(146
97)

\textref h:C v:C \htext(10
90){\framebox{\scriptsize$\def\arraystretch{0.8}\begin{array}{c}
S^{(2,1)}\\ {[}\left\langle C_{2x}{\cal T},C_{3z}{\rm
U}_1(\frac{4\pi}{3})\right\rangle {]} \end{array}$}}

\textref h:C v:C \htext(47
90){\framebox{\scriptsize$\def\arraystretch{0.8}\begin{array}{c}
S^{(2,2)}\\ {[}\left\langle C_{2x}{\cal T},C_{2z}{\rm
U}_1(\pi)\right\rangle {]}  \end{array}$}}

\textref h:C v:C \htext(79
90){\framebox{\scriptsize$\def\arraystretch{0.8}\begin{array}{c}S^{(2,3)}\\
{[}\left\langle C_{2x},C_{4z}{\cal T} \right\rangle
{]}\end{array}$}}

\textref h:C v:C \htext(111
90){\framebox{\scriptsize$\def\arraystretch{0.8}\begin{array}{c}
S^{(2,4)}\\ {[}\left\langle C_{2x},C_{2z},{\cal T}\right\rangle
{]}\end{array}$}}

\textref h:C v:C \htext(146
90){\framebox{\scriptsize$\def\arraystretch{0.8}\begin{array}{c}
S^{(2,5)}\\ {[}\left\langle C_{4z}{\rm U}_1(\pi),C_{2x}{\cal
T}\right\rangle {]} \end{array}$}}

\move(10 83) \avec(47 37) \move(47 83) \avec(47 37) \move(79 83)
\avec(79 67) \move(79 83)  \avec(111 67) \move(111 83) \avec(79
67) \move(111 83)  \avec(111 67) \move(146 83) \avec(111 67)

\textref h:C v:C \htext(79
60){\scriptsize\framebox{$\def\arraystretch{0.8}\begin{array}{c}
S^{(3,2)}\\ {[}\left\langle C_{2x}C_{2z}\right\rangle {]}
\end{array}$}}

\textref h:C v:C \htext(111
60){\scriptsize\framebox{$\def\arraystretch{0.8}\begin{array}{c}
S^{(3,1)}\\ {[}\left\langle C_{2z},C_{2x}{\cal T}\right\rangle
{]}\end{array}$}}

\move(79 54) \avec(79 37) \move(111 54) \avec(47 37) \move(111 54)
\avec(79 37)

\textref h:C v:C \htext(47
30){\scriptsize\framebox{$\def\arraystretch{0.8}\begin{array}{c}
S^{(4,1)}\\ {[}\left\langle
 C_{2x}\,{\cal T}\right\rangle {]}\end{array}$}}

\textref h:C v:C \htext(79
30){\scriptsize\framebox{$\def\arraystretch{0.8}\begin{array}{c}
S^{(4,2)}\\ {[}\left\langle C_{2z} \right\rangle {]}
\end{array}$}}

\move(46 24) \avec(80 7) \move(79 24) \avec(80 7)

\textref h:C v:C \htext(80
0){\scriptsize\framebox{$\def\arraystretch{0.8}\begin{array}{c}
S^{(6,1)}=S_{\rm principal} \\
 \{{\bf e}\}\end{array}$}}

\end{texdraw}
\end{figure}
\end{center}

\begin{center}
\begin{figure}
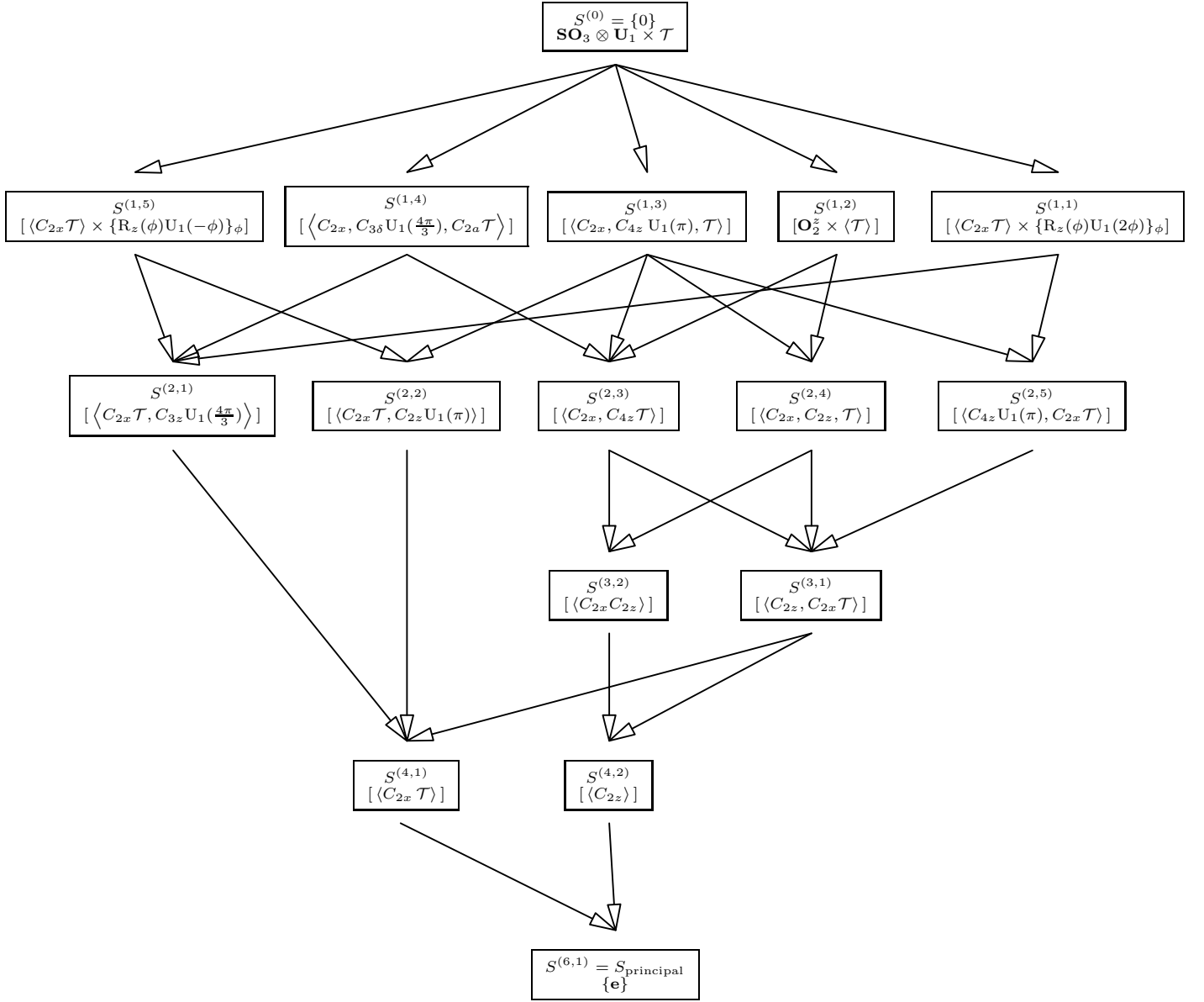

\caption{Localization of the absolute minimum of a fourth degree $G$-invariant polynomial,
$\widehat\Phi^{(4)}(\alpha,p)=\alpha_0 p_1^2/2 + \sum_{j=1}^3 \alpha_j p_j$, $\alpha_1<0$, as a function of its
coefficients. For values of $(\alpha_0,\alpha_2,\alpha_3)$ in ${\cal R}_1$ or in ${\cal R}_2$ or in ${\cal R}_3$, the
absolute minimum lies, respectively, in
the strata $S^{(1,4)}$ or $\{S^{(1,2)},S^{(1,3)},S^{(2,4)}\}$ (degenerate minimum) or $S^{(1,1)}$. For particular values of
the $\alpha$'s, see Table~\ref{T4}.}

\begin{texdraw}

\drawdim mm

\setunitscale 0.6

\move(100 40) \avec(100 180) \move(34 100) \avec(180 100)
\textref h:C v:C \htext(95 171){$\small\alpha_2$}
\textref h:C v:C \htext(171 95){$\small\alpha_3$}

\move(100 100) \linewd 0.8 \lvec(90 160)
\move(100 100) \linewd 0.8 \lvec(165 100)
\move(100 100) \linewd 0.8 \lvec(70 40)

\textref h:C v:C \htext(60 105){${\cal R}_2$:\ ${\small\alpha_0 > -2\,\alpha_2/3}$}
\textref h:C v:C \htext(140 130){${\cal R}_1:\ {\small\alpha_0 > -2\,\alpha_2}$}
\textref h:C v:C \htext(136 70){${\cal R}_3:\ {\small\alpha_0 > -\alpha_2-2\,\alpha_3}$}

\textref h:C v:C \rtext td:280 (90 140){$\scriptstyle\alpha_2=-6\,\alpha_3$}
\textref h:C v:C \rtext td:65 (74 55){$\scriptstyle\alpha_2=2\,\alpha_3$}
\textref h:C v:C \htext(135 103){$\scriptstyle\alpha_2=0$}

\setgray 1 \lvec(100 195)

\end{texdraw}
\end{figure}
\end{center}

 
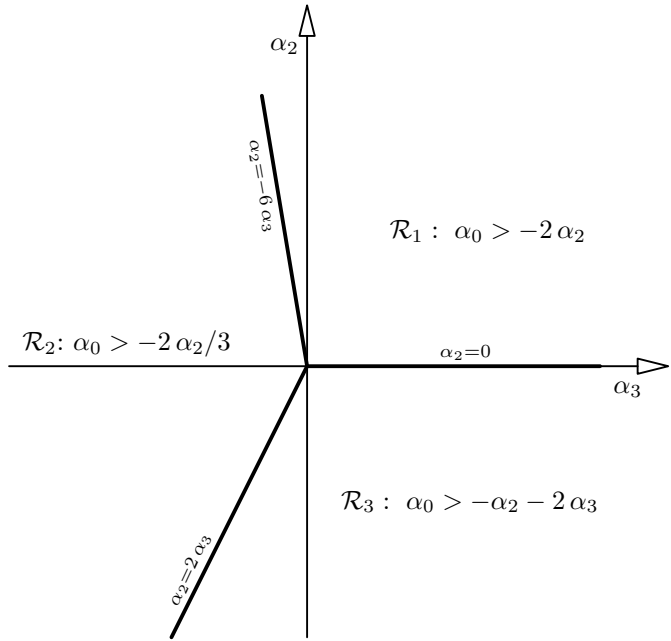
\end{document}